\documentstyle[11pt]{article}
\include{epsf}
\begin{document}

\thispagestyle{empty}

\begin{flushright}
UNIGRAZ-UTP-28-06-98 \\
hep-lat/9806032
\end{flushright}
\begin{center}
\vspace*{5mm}
{\Large Clover improvement, spectrum and Atiyah-Singer
\vskip2mm index theorem for the Dirac operator on the lattice$^*$}
\vskip9mm
\centerline{ {\bf
Christof Gattringer}}
\vskip 2mm
\centerline{Department of Physics and Astronomy,}
\centerline{University of British Columbia, Vancouver B.C., Canada}
\vskip 5mm
\centerline{ {\bf
Ivan Hip}}
\vskip2mm
\centerline{Institut f\"{u}r Theoretische Physik} 
\centerline{Universit\"at Graz, A-8010 Graz, Austria}
\vskip10mm
\begin{abstract}
We study the role of the $O(a)$-improving clover term for the spectrum of 
the lattice Dirac operator using cooled and thermalized SU(2)
gauge field configurations.
For cooled configurations we observe improvement of the spectral properties
when adding the clover term. For the thermalized case 
($12^4, \beta = 2.4$) without clover term we 
find a rather bad separation of physical and doubler branches making a 
probabilistic interpretation of the Atiyah-Singer index theorem on the
lattice questionable for this $\beta$ and lattice size. 
Adding the clover term leads to the creation of additional
real eigenvalues which come in pairs of opposite chirality thus further 
worsening the situation for the index theorem.
\end{abstract}
\end{center}
\vskip8mm \nopagebreak \begin{flushleft} \rule{2 in}{0.03cm}
\\ {\footnotesize \ 
${}^*$ Supported by Fonds zur F\"orderung der Wissenschaftlichen 
Forschung in \"Osterreich, Projects P11502-PHY and J1577-PHY.}
\end{flushleft}
\newpage
\setcounter{page}{1}
\noindent
{\bf 1. Introduction}
\\
In the last few years the spectrum of the Dirac operator on the lattice 
has seen a lot of attention. This is partly motivated by the hope that 
a lattice-regularized gauge theory could put the idea of decomposing 
the fully quantized path integral into topological sectors on a 
conceptually sound basis. 
In the continuum gauge fields can be classified with respect to their 
topological charge when they are smooth and obey certain boundary 
conditions. On the other hand the fields which carry the measure in 
a continuum path integral do not obey these conditions. The topological 
arguments thus can only be implemented on a semi-classical level. 
Lattice regularization might provide the framework to set up the 
topological concepts in a fully quantized theory. Up to so-called
exceptional configurations lattice gauge fields can be assigned a 
topological charge and the exceptional configurations are expected 
to die out in the continuum limit. A complete decomposition of the 
path integral into topological sectors could become possible. 

In the continuum the Atiyah-Singer index theorem \cite{AtSi68} relates 
the topological charge of classical gauge field configurations 
to the numbers of left- and right-handed zero modes of the Dirac operator.
This gives rise to semi-classical results for fermionic observables.
Again, these results can not be implemented in the fully quantized 
model since the gauge fields in the path integral are too rough. 

On the lattice there is no analytic result for the Atiyah-Singer 
index theorem, but it has been conjectured
\cite{ItIwYo87}-\cite{SmSiTe98} that on the lattice, at least 
for sufficiently smooth configurations, it is realized in a probabilistic
sense. The question however arises if for the parameters where current 
simulations are done the gauge fields are sufficiently smooth so that 
the spectrum of the Dirac operator can be interpreted in the sense 
of the index theorem. 

For completely smooth configurations such as lattice approximations to 
continuum instantons there can be no doubt, that the 
topological interpretation of the spectrum makes sense 
\cite{smooth}-\cite{EdHeNaSi98}.
Also for fully quantized QED$_2$ it has been established
\cite{qed2}-\cite{GaHiLa97}, that the 
index theorem governs the behavior of the spectrum when approaching
the continuum limit. 

For thermalized lattice gauge theories in four 
dimensions the picture is less conclusive. Older results by 
Itoh, Iwasaki and Yoshi\'e \cite{ItIwYo87} report
a rather poor manifestation of the index theorem and also the distinction 
of physical and doubler modes turned out to be difficult.
More recent studies \cite{NaVr97,NaSi98} 
(compare also \cite{Ne97},\cite{Naetal98}-\cite{SmSiTe98} 
for results on the spectrum for thermalized gauge fields)
of a fermionic definition of the topological charge,
which is based on a lattice manifestation of the index theorem, are more 
optimistic. However, only sampled data but no detailed analysis of the 
index theorem for single thermalized configurations was presented. 

One of the motivations for this contribution 
is to carefully analyze the idea of interpreting the spectrum 
of the lattice Dirac operator in terms of a probabilistic manifestation
of the Atiyah-Singer index theorem.
This is done for a sample of thermalized SU(2) configurations 
on $12^4$ lattices at $\beta = 2.4$ taken from the quenched 
simulation \cite{FoGaSt97} (sample 1b).
The authors of \cite{FoGaSt97} were so kind to provide us also with 
their cooled data obtained using the improved cooling method
(compare also \cite{impcool2})
based on the over-improved action given in \cite{GaGoSnBa94}. 
This allows to compare the spectra for the thermalized
configurations with the cooled counterparts on a one by one basis.
One can check if a fermionic definition of the topological 
charge based on the index theorem, if possible
at all, agrees with the cooling procedure. Our results 
presented here indicate however,
that for SU(2) quenched configurations on $12^4$ lattices at $\beta = 2.4$
most of the configurations give rise to a spectrum where no
clear separation of physical and doubler modes can be established.
An interpretation of the spectrum in terms of the index theorem is
questionable.

In the last few years it was realized that improvement is a powerful 
if not essential tool when numerically analyzing lattice field theories
(see e.g.~\cite{Lu98} for an introductory overview). For the fermion action 
the $O(a)$-improvement is governed by the clover term. Adding this extra
term to the lattice Dirac operator certainly alters its spectral
properties. 

The second motivation for this study of the eigensystem of the 
lattice Dirac operator is to see how the clover term changes the 
spectrum. Does it improve the spectral properties similar to the 
behavior which was observed in QED$_2$ \cite{qed2perf} using
perfect actions \cite{perfact}? Again we address this question by 
analyzing spectra for the thermalized, quenched sample on $12^4$ 
lattices at $\beta = 2.4$ and also for the corresponding cooled 
configurations. In addition we analyze the effect of the clover 
term for complete spectra using smooth toy configurations
(constant plaquette configurations + small
fluctuations) on $4^4$ lattices.

The article is organized as follows: In the next section we will set our 
notation and briefly discuss the framework for the Atiyah-Singer index
theorem on the lattice. Section 3 presents our results for the effect
of the clover term on complete spectra using toy configurations on $4^4$
lattices. Section 4 contains the study of spectra for cooled configurations
on $12^4$ lattices. The index theorem and the effects of the 
clover term are discussed.
In Section 5 we present our results for thermalized configurations on 
$12^4$ lattices at $\beta = 2.4$. We analyze in detail the mechanism for
the proliferation of real eigenvalues caused by the clover term using 
perturbative arguments. In Section 6 we assess the possibility for the 
interpretation of the spectrum in the sense of the index theorem. 
The article closes with a discussion. 
\\
\\
{\bf 2. Notation and basic properties of the lattice Dirac operator}
\\
We consider the following, $O(a)$-improved fermion action
\begin{equation}
S = \sum_{x,y} \overline{\psi}(x) D(x,y) \psi(y) \; , 
\label{fmatrix}
\end{equation}
where the lattice Dirac operator $D$ (fermion matrix) consists of three parts
\begin{equation}
D \; = \; (4 + m) \mbox{1\hspace{-1.1mm}I} \; - \; K \; \; \; \; \; \; ,
\; \; \; \; \; K \; = \; Q \; - \; c_{sw} C \; ,
\end{equation}
with $m$ denoting the bare quark mass. For later convenience we 
introduced the abbreviation $K$ for the non-trivial part of the 
fermion matrix. The Wilson hopping matrix $Q$ and the 
clover term $C$ are given by
\begin{eqnarray}
Q(x,y)\!&\!=\!&\!\frac{1}{2} \sum_{\mu} \Big\{
[ 1 - \gamma_\mu ] U_\mu(x) \delta_{x+\mu, y}
+ [ 1 + \gamma_\mu ] U_\mu(x-\mu)^\dagger 
\delta_{x-\mu, y} \Big\}, \\
C(x,y) &\!=\!& \sum_{\mu,\nu} \frac{i}{4} \sigma_{\mu\nu} 
F_{\mu \nu}(x) \; \delta_{x,y} \; . 
\label{cloverterm}
\end{eqnarray}
Here $\sigma_{\mu \nu} = i/2 [\gamma_\mu, \gamma_\nu]$ and $F_{\mu \nu}$
is some lattice discretization of the continuum field strength tensor.
We use the standard (clover) discretization which can e.g.~be found in 
\cite{LuSiSoWe97}. 
For a proper choice of the Sheikholeslami-Wohlert 
coefficient $c_{sw}$ in (\ref{fmatrix}), the 
counter-term (\ref{cloverterm}), first given in \cite{ShWo85}, removes 
$O(a$) cutoff effects. For SU(2) the one loop perturbative expansion 
gives \cite{ShWo85,LuWe96}
\begin{equation}
c_{sw} \; = \; 1 \; + \; 0.155(1) \; g_0^2 \; + \; O(g_0^4) \; ,
\label{cpert}
\end{equation}
where $g_0$ is the bare coupling, related to $\beta$ used below via
$\beta = 4/g_0^2$. As in the case without improvement, the matrix 
$K$ (or equivalently $D$) is 
neither hermitian nor anti-hermitian, but hermitian conjugation 
is implemented as a similarity transformation  
\begin{equation}
\gamma_5 K \gamma_5 \; = \; K^\dagger \; .
\label{hercon}
\end{equation}
This equation implies that the eigenvalues of $K$ (or $D$)
come in complex conjugate pairs or are real.
The staggered transformation which transforms $Q$ into $-Q$ 
(see e.g.~\cite{MoMu94}) leaves the clover term $C$ invariant. Thus for 
$c_{sw} \neq 0$ the symmetry of the spectrum of $K$ under reflection 
at the imaginary axis is lost.

Besides leading to symmetry of the spectrum under complex conjugation,
(\ref{hercon}) also has an interesting implication \cite{ItIwYo87}
for the chiral properties of the eigenvectors of $K$: 
Only eigenvectors $\psi_\lambda$ of $K$ ($D$), where the eigenvalue $\lambda$ is
real can have non-vanishing pseudoscalar matrix elements
\begin{equation}
(\psi_\lambda, \gamma_5 \psi_\lambda) \; \neq \; 0 \; \; \; 
\mbox{only for} \; \; \; \lambda \in \mbox{I\hspace{-1.0mm}R} \; \; .
\label{chiform}
\end{equation}
For the Dirac operator in the continuum 
the only eigenstates with non-vanishing
pseudoscalar matrix elements are zero modes. 
Thus the eigenvectors 
$\psi_\lambda$ of the non-hermitian 
matrix $K$ (or $D$) which have real eigenvalues 
$\lambda$ should be interpreted as 
the "lattice zero modes" \cite{ItIwYo87,SmVi87,NaNe95}.

Based on this interpretation of the eigenvectors with real 
eigenvalues as the lattice zero modes, one can formulate a 
lattice version of the Atiyah-Singer index theorem \cite{AtSi68} 
which is expected to hold for sufficiently smooth gauge fields
\begin{equation}
\nu[U] \; = \; R_- \; - \;  R_+ \; .
\label{lasit}
\end{equation} 
Here $R_+$ and $R_-$ are the numbers of real 
eigenvalues in the physical branch of the spectrum with 
positive and negative chirality. The chirality is defined 
as the sign of the pseudoscalar matrix element $(\psi, \gamma_5 \psi)$ 
of the corresponding eigenvectors $\psi$. $\nu[U]$ denotes the 
topological charge of the lattice gauge field configuration.
\\
\\
{\bf 3. Complete spectra for test configurations on small lattices}
\\
Relatively few is known about the role of the clover term for 
the spectrum of the lattice Dirac operator 
(for some results see \cite{Naetal98,BaDuEiHoTh98,SmSiTe98}).
To gain some basic insight on the role of the clover term $C$ for all 
eigenvalues of $K$ we computed several complete spectra of $K$ on 
$4^4$ lattices. These computations were also performed to test the correct 
implementation of the fermion matrix by checking the linear and 
quadratic sum rules for the eigenvalues which are obtained by expressing 
the trace of $K$ and $K^2$ as a functional of the gauge field. Both sum 
rules are obeyed with excellent numerical accuracy. The computations in 
this section were done using standard routines for general 
complex matrices from the LAPACK package. 

\begin{figure}[htp]
\centerline{\hspace*{-2mm}
\epsfysize=5.3cm\epsfbox[62 72 508 434] {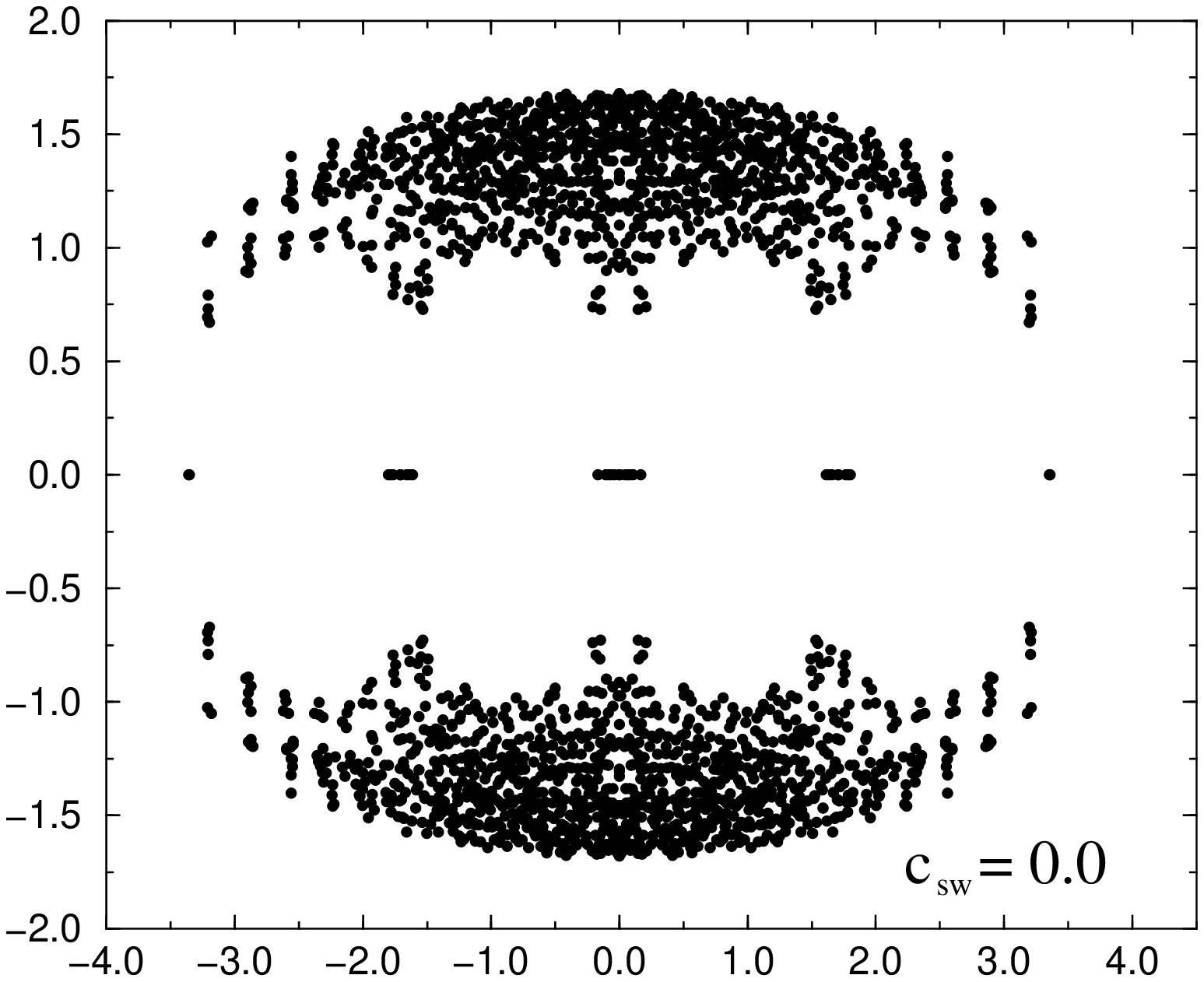}
\hspace{-3mm}
\epsfysize=5.3cm\epsfbox[87 72 508 434] {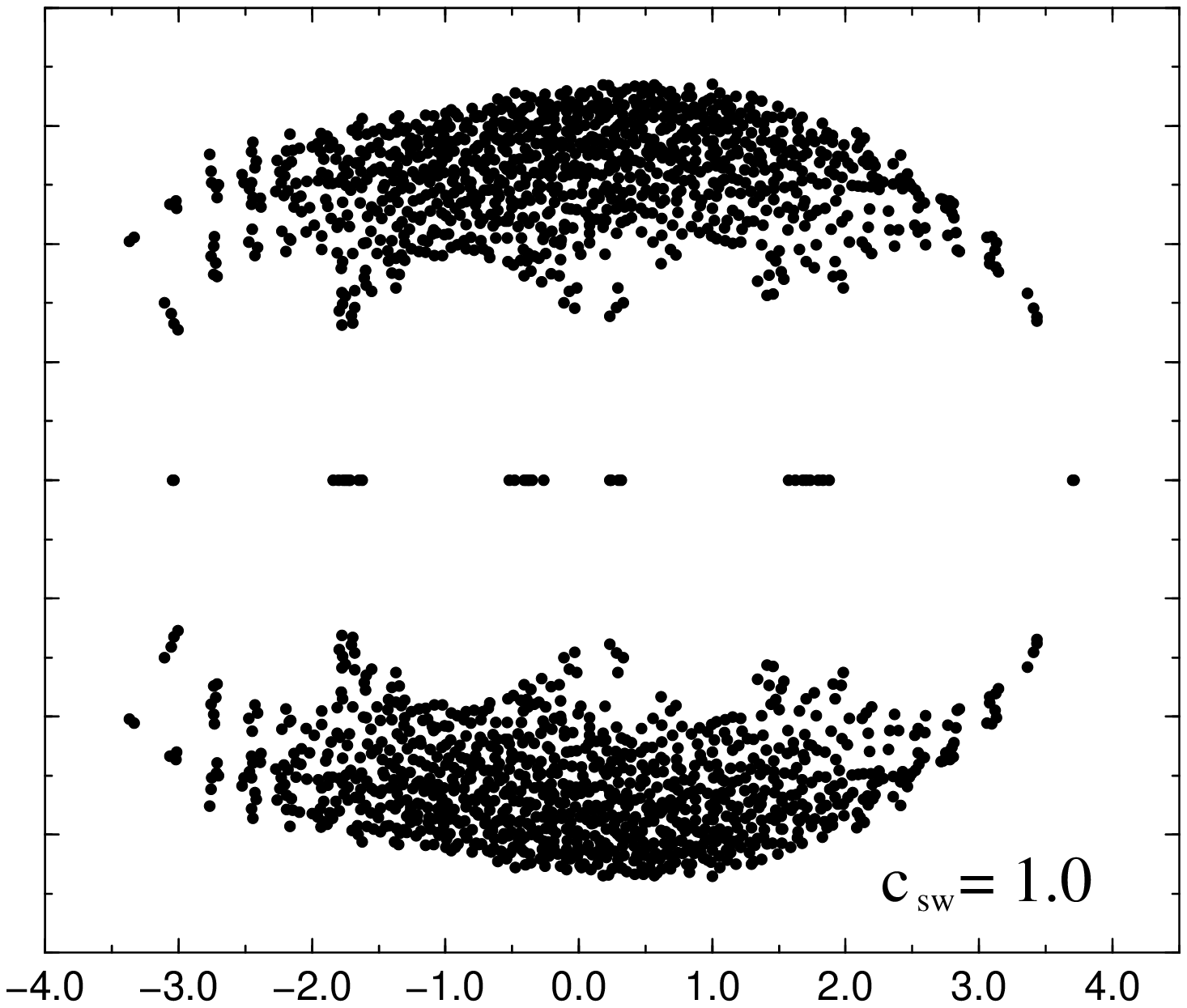} 
} 
\vspace{3mm}
\centerline{\hspace*{-2mm}
\epsfysize=5.3cm\epsfbox[62 72 508 434] {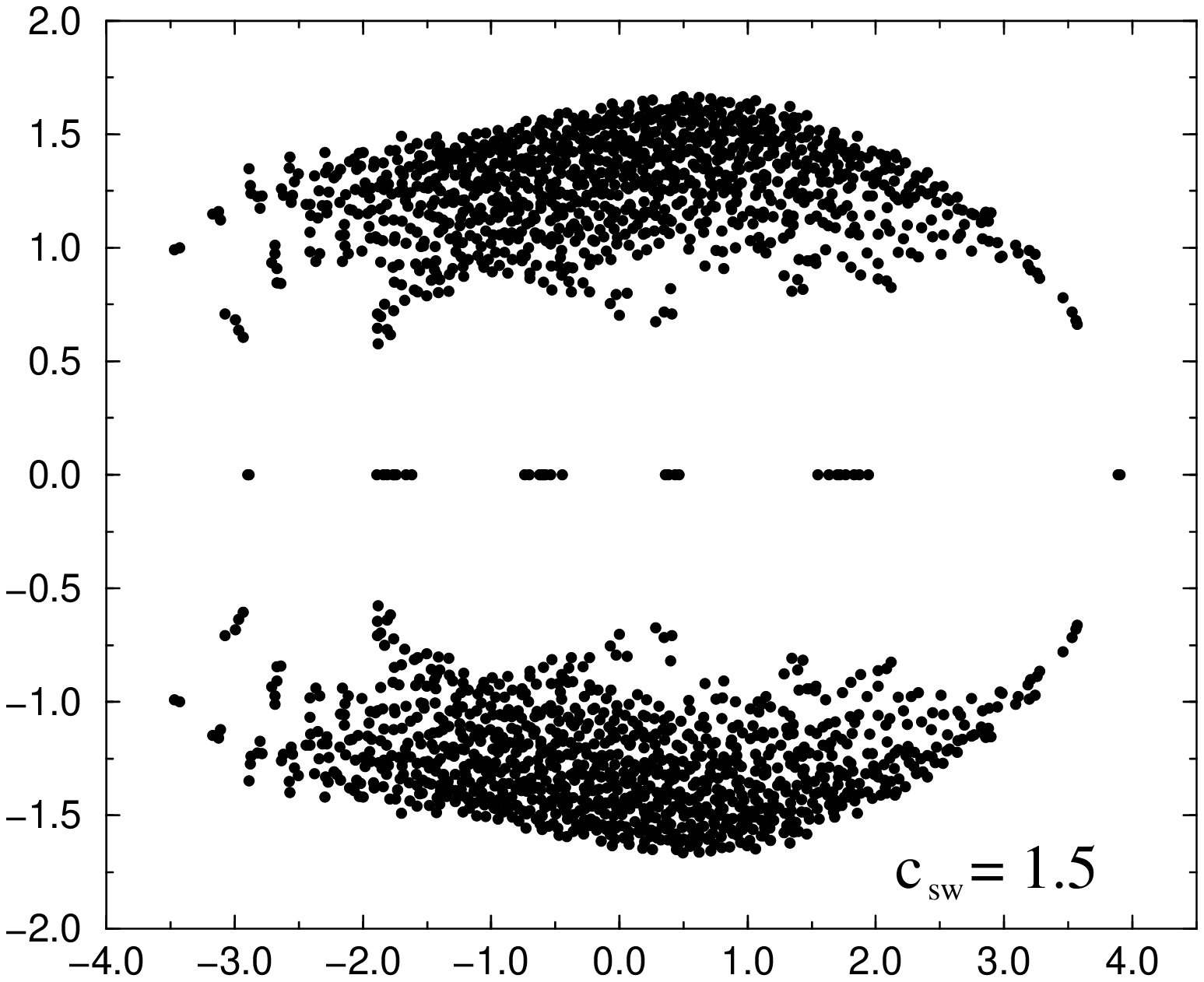}
\hspace{-3mm}
\epsfysize=5.3cm\epsfbox[87 72 508 434] {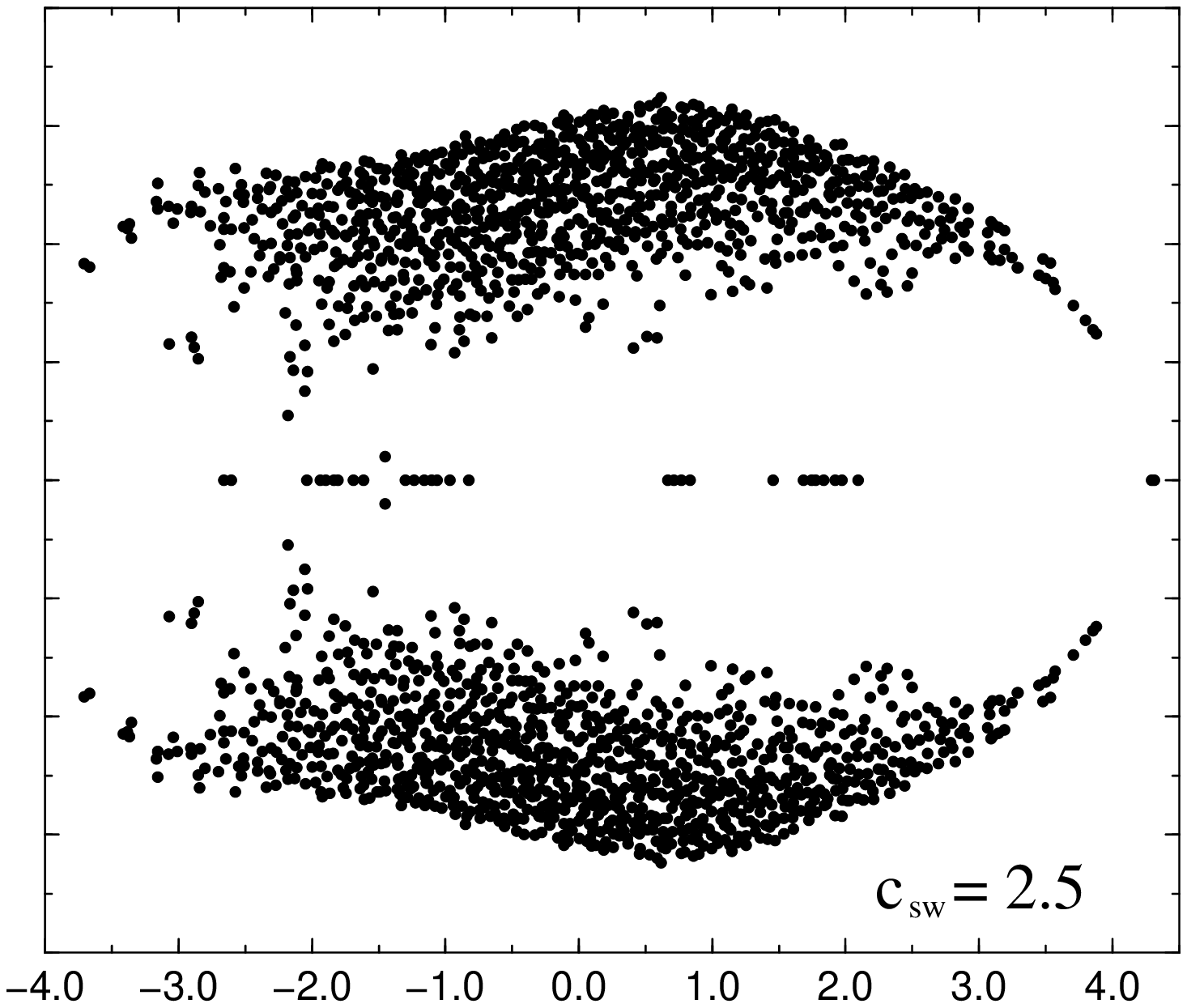} 
}
\caption{{\sl Dependence of the complete spectrum of $K$ on $c_{sw}$. 
We plot all 
eigenvalues of $K$ in the complex plane
for a relatively smooth gauge field with topological charge 
$\nu[U] = 2$. 
The coefficient $c_{sw}$ was chosen to be $c_{sw} = 0.0, 1.0, 1.5$ 
and $2.5$.
There are two real eigenvalues with negative chirality in the physical 
branch (in the vicinity of $4$ on the real axis), in accordance with} 
(\protect{\ref{lasit}}).
\label{spectra}}
\end{figure}

The gauge fields were constructed from constant field strength 
configurations where we added some fluctuations. For the details of 
the preparation of 
the gauge field see \cite{GaHi97}. In Fig.~\ref{spectra} we show 
spectra of $K$ for different values of $c_{sw}$. The background field $U$
has topological charge $\nu[U] = 2$ and is relatively smooth (in the 
notation of \cite{GaHi97}: $s=t=1,\; \varepsilon = 0.1,\; n_r = 15$). 
We show the spectra for values of the Sheikholeslami-Wohlert coefficient 
$c_{sw} = 0.0, 1.0, 1.5$ and 2.5. 
We remark, that the gauge fields used for this
computation are static background fields which cannot be assigned an
inverse squared coupling $\beta$ 
and here the perturbative formula (\ref{cpert}) can
only give an idea about the order of magnitude of $c_{sw}$. 

The first plot in Fig.~\ref{spectra} shows the spectrum of the 
un-improved ($c_{sw} = 0$) operator $K$ in the complex plane. 
It displays the 
well known features of the spectrum for a rather smooth
background gauge field with topological charge $\nu = 2$. 
The spectrum is symmetric
with respect to reflection at real and imaginary axis. There is a bulk of
complex eigenvalues, well separated from the real axis and several 
real eigenvalues clustering in small regions, corresponding to the
physical and doubler branches. 
The physical branch (right edge of the spectrum) is well pronounced 
for this rather smooth gauge field configuration.
There are no problems distinguishing the real eigenvalues in the physical 
branch of the spectrum from the doublers. The two real eigenvalues in the 
physical branch have negative chirality in accordance with (\ref{lasit}).

When the Sheikholeslami-Wohlert coefficient $c_{sw}$ is chosen equal to 1
the spectrum changes. The symmetry under reflection at the imaginary 
axis is lost, since the clover term $C$ is even under the staggered 
transformation, while the standard hopping term $Q$ is odd (compare the
discussion above). However the symmetry under reflection at the real axis,
which is due to (\ref{hercon}) remains. 
The whole physical branch is shifted 
towards larger real parts (compare also
the perturbative analysis in Section 5), and the complex eigenvalues in the 
physical branch seem to cluster more along a single curve. 
The discussed deformation of the spectrum becomes more pronounced when 
increasing $c_{sw}$ to $c_{sw} = 1.5$ and 2.5.  

It is remarkable, that for this rather smooth gauge field configuration 
the two real eigenmodes, relevant for the index theorem, 
remain rather unperturbed when increasing $c_{sw}$.
We will later demonstrate that for thermalized configurations the changes
of the spectrum caused by the clover term are much more dramatic.

We end this section with remarking, that from the last plot it is 
obvious that the bound known at $c_{sw} = 0$ for the 
eigenvalues $\lambda$ of $K$: 
$|\lambda| < 4$, or more general $|\lambda| < D$ where $D$ is 
the dimension, does not apply at $c_{sw} > 0$. 
The standard strategy \cite{WeCh79} for proving 
this bound cannot be implemented for $c_{sw} \neq 0$, and in fact the 
plot shows that the bound is violated already for not too large $c_{sw}$.
\\
\\
{\bf 4. The physical branch of the spectrum - cooled configurations}
\\
In this section we study in more detail the role of the improvement 
term for the physical branch of the spectrum in smooth gauge field 
configurations. In particular we use cooled configurations obtained from 
thermalized configurations from a quenched simulation of SU(2) gauge 
theory \cite{FoGaSt97}. 

\begin{figure}[htp]
\centerline{\hspace*{-1mm}
\epsfysize=7cm\epsfbox[350 72 509 434] {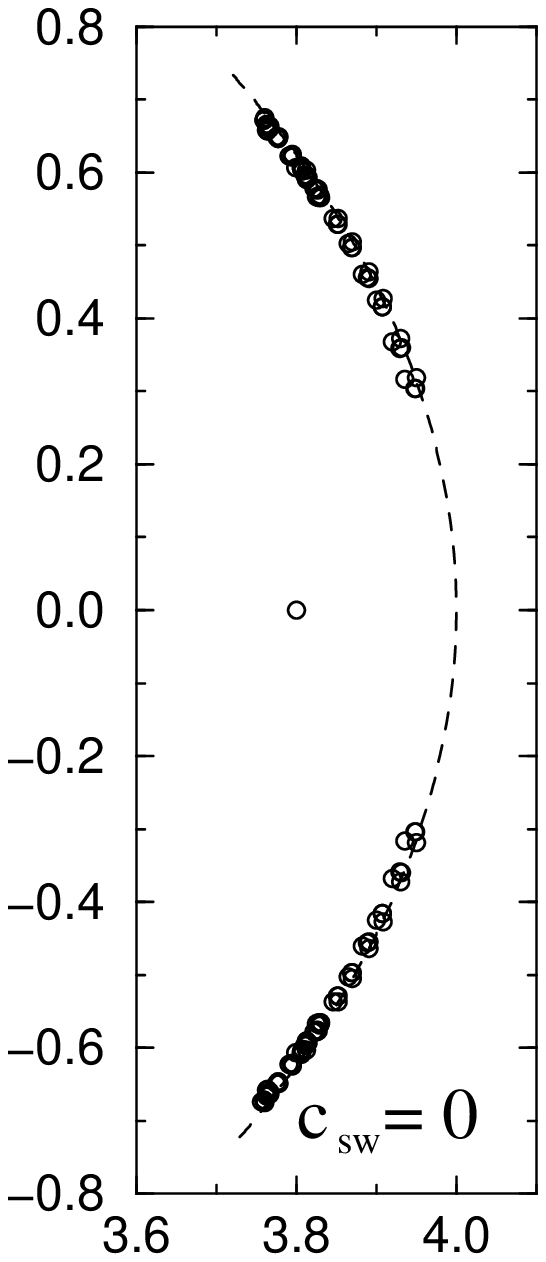}
\hspace{-3mm}
\epsfysize=7cm\epsfbox[381 72 509 434] {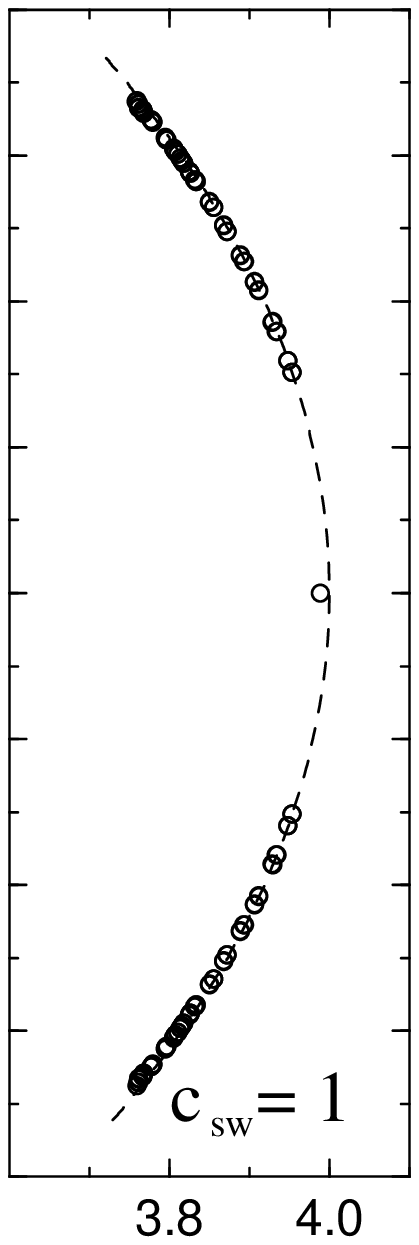} 
\hspace{8mm}
\epsfysize=7cm\epsfbox[350 72 509 434] {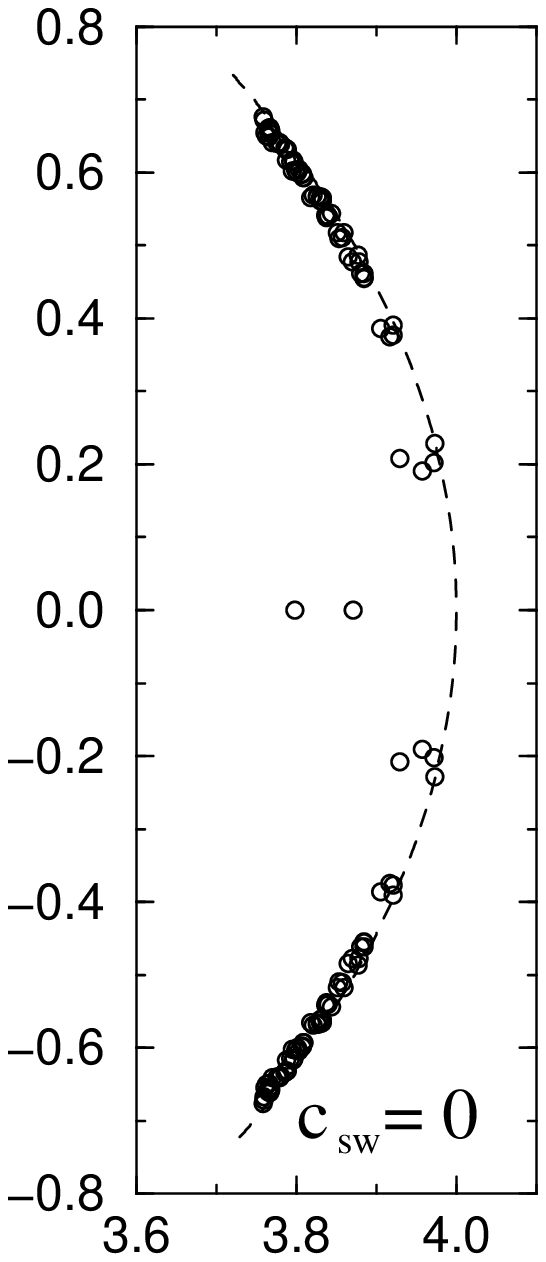}
\hspace{-3mm}
\epsfysize=7cm\epsfbox[381 72 509 434] {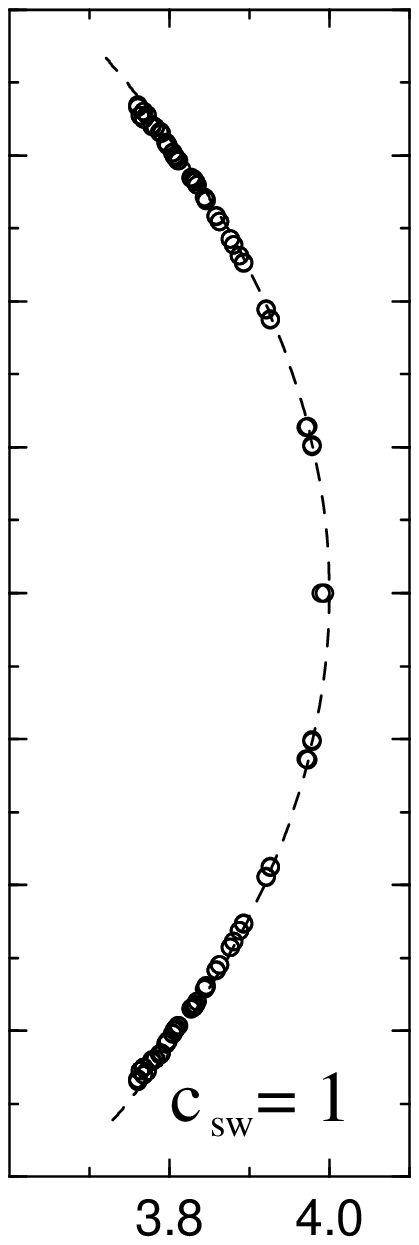} 
}
\caption{{\sl
Plots of $128$ eigenvalues in the physical branch of the spectrum of $K$
for cooled gauge field configurations with $\nu[U] = -1$ 
(left hand side plot)
and $\nu[U] = 2$ (right hand side plot). 
For both configurations we show the spectra for
$c_{sw} = 0$ (no improvement) and
$c_{sw} = 1$ (perturbative value for smooth fields). The dashed curve is 
the ellipse centered at the origin of the complex plane
with width $8$ and height $4$. We remark, that in the 
right-most plot the two real eigenvalues are very close to each other, 
such that the two symbols appear as one.}
\label{smoothplots}}
\end{figure}

To analyze the physical branch of the spectrum we use the 
Implicitly Restarted Arnoldi/Lanczos
Method \cite{So92} which allows to compute a few eigenvalues with 
user specified properties for large, general matrices. 
We set up the algorithm such that the 128 eigenvalues 
with the largest real parts are
computed first. This gives the eigenvalues in the physical 
branch of the spectrum we are interested in.
In Fig.~\ref{smoothplots} we compare the physical branch of the 
spectrum of $K$ for $c_{sw} = 0$ and 
$c_{sw} = 1$ for two cooled configurations from \cite{FoGaSt97}, 
one with $\nu[U] = -1$ (left hand side plot) and one with $\nu[U] = 2$
(right hand side plot). The value $c_{sw} = 0$  
is the case without $O(a)$-improvement
and $c_{sw} = 1$ is the perturbative result 
(\ref{cpert}) for smooth configurations ($g_0 = 0$).

The plots show, that already without improvement
the complex eigenvalues in the
physical branch are concentrated in a small band along the ellipse with 
width 8 and height 4 centered at the origin (dashed curve in
Fig.~\ref{smoothplots}). When adding the improvement term this alignment 
along the ellipse is further enhanced.

For the real eigenvalues the situation is different. 
Without improvement they 
are considerably shifted away from the limiting curve, 
while the clover term 
brings them very close to the ellipse, i.e.~close to the origin 
when looking at the spectrum of $D$. Thus the real eigenvalues 
become exact poles of the improved massless propagator in smooth
background fields. The number of real eigenvalues 
and their chirality is in accordance with (\ref{lasit}). 
The value of the pseudoscalar matrix element is 
considerably improved by the clover term, i.e.~much closer to $\pm 1$
(compare also \cite{SmSiTe98}). As an example we give the values for
the $\nu = -1$ configuration (left hand side plot) where we obtain 0.8009
for the matrix element without improvement and 0.9984 at $c_{sw} = 1$. 

We analyzed several more cooled configurations from sample 1b of 
\cite{FoGaSt97} and found the above results for cooled configurations 
confirmed. The number of real eigenvalues in the physical branch and 
their chirality are properly described by 
(\ref{lasit}) both for $c_{sw} = 0$
and $c_{sw} = 1$. The clover term improves the alignment of the complex 
eigenvalues along the ellipse, shifts the real eigenvalues and leads to
pseudoscalar matrix elements very close to $\pm 1$. This improvement of the
spectral properties for smooth configurations when adding the clover term
is somewhat surprising for a term which was designed for improving on-shell
quantities. It is a little bit reminiscent of the results obtained for the 
spectrum of the perfect lattice Dirac operator in QED$_2$ \cite{qed2perf}.
\\
\\
{\bf 5. Thermalized configurations}
\\
The literature \cite{smooth}-\cite{EdHeNaSi98}, \cite{SmSiTe98}
as well as the above results for the cooled configurations 
and the test configurations in Section 3 show, that the lattice 
index theorem 
(\ref{lasit}) holds perfectly well for the spectrum in smooth gauge field 
configurations. For the case of thermalized configurations the situation 
is less clear \cite{ItIwYo87},\cite{NaVr97}-\cite{SmSiTe98}. 

In this section we carefully analyze the role of $c_{sw}$ for
the spectrum for thermalized configurations from sample 1b in 
\cite{FoGaSt97}, i.e. quenched SU(2) gauge field configurations at 
$\beta = 2.4$ on a $12^4$ lattice.
We compare these results to the results for the cooled 
counterparts obtained in the last section.  

Again we use the Implicitly Restarted Arnoldi/Lanczos
Method to compute 200 complex eigenvalues with largest real part, i.e.
eigenvalues in the physical branch of the spectrum. In addition we 
compute eigenvalues on the real axis using the method of counting level
crossings of the hermitian modification of the fermion matrix. This method
was first described in \cite{ItIwYo87} and brought to further perfection 
and used for analyzing the real spectrum of the Dirac operator 
\cite{NaNe95,EdHeNaSi98,NaVr97,NaSi98,Naetal98,SmSiTe98}. It is now
also known as "overlap method".

Whenever $K$ has a real eigenvalue $r$, then the operator 
$\rho$1\hspace{-1mm}I $ - K$ has a zero eigenvalue at $\rho = r$. 
This implies furthermore \cite{ItIwYo87}, that the auxiliary matrix 
\begin{equation}
H(\rho) \; \equiv \; \gamma_5 [ \; 
\rho \mbox{1\hspace{-1mm}I} \; - \; K \; ] \;,
\label{auxi}
\end{equation}
has a zero eigenvalue at $\rho = r$ 
if and only if $r$ is a real eigenvalue of $K$. Thus one can 
trace the flow of small eigenvalues $\mu(\rho)$ of $H(\rho)$ as $\rho$ 
is varied. Whenever the flow crosses zero at some $\rho = r$, then this 
$r$ is a real eigenvalue of $K$. From the definition (\ref{auxi})
for the auxiliary operator $H(\rho)$ it is also obvious that 
the pseudoscalar matrix element corresponding to $r$ is 
given by the slope of the
flow for the eigenvalue $\mu(\rho)$ which crosses 
zero at $r$ \cite{ItIwYo87}
\begin{equation}
(\psi_r, \gamma_5 \psi_r) \; = 
\; \frac{d}{d\rho} \mu(\rho) \Bigg|_{\rho \; = \; r} \; .
\label{slope}
\end{equation}
Thus the chirality assigned to $r$ 
is given by the sign of the slope. The auxiliary operator
$H(\rho)$ is hermitian due 
to (\ref{hercon}) and thus much simpler to deal with numerically.
A more detailed discussion and 
technical remarks on the implementation of the level crossing 
algorithm can be found in the literature 
\cite{ItIwYo87,NaNe95,EdHeNaSi98,NaVr97,NaSi98,Naetal98,SmSiTe98}.

Here we search for the real spectrum between 1.8 and 4. 
Real eigenvalues in the
vicinity of 2 are associated with the 4 doubler modes from the 
$(\pi,0,0,0)$-type corners of the Brillouin zone. 
In order to make sure that we find all real eigenvalues in the physical 
branch and at least some of the doublers we start our search at 1.8.

\begin{figure}[htp]
\centerline{\hspace*{-1mm}
\epsfysize=5.3cm\epsfbox[56 68 519 434] {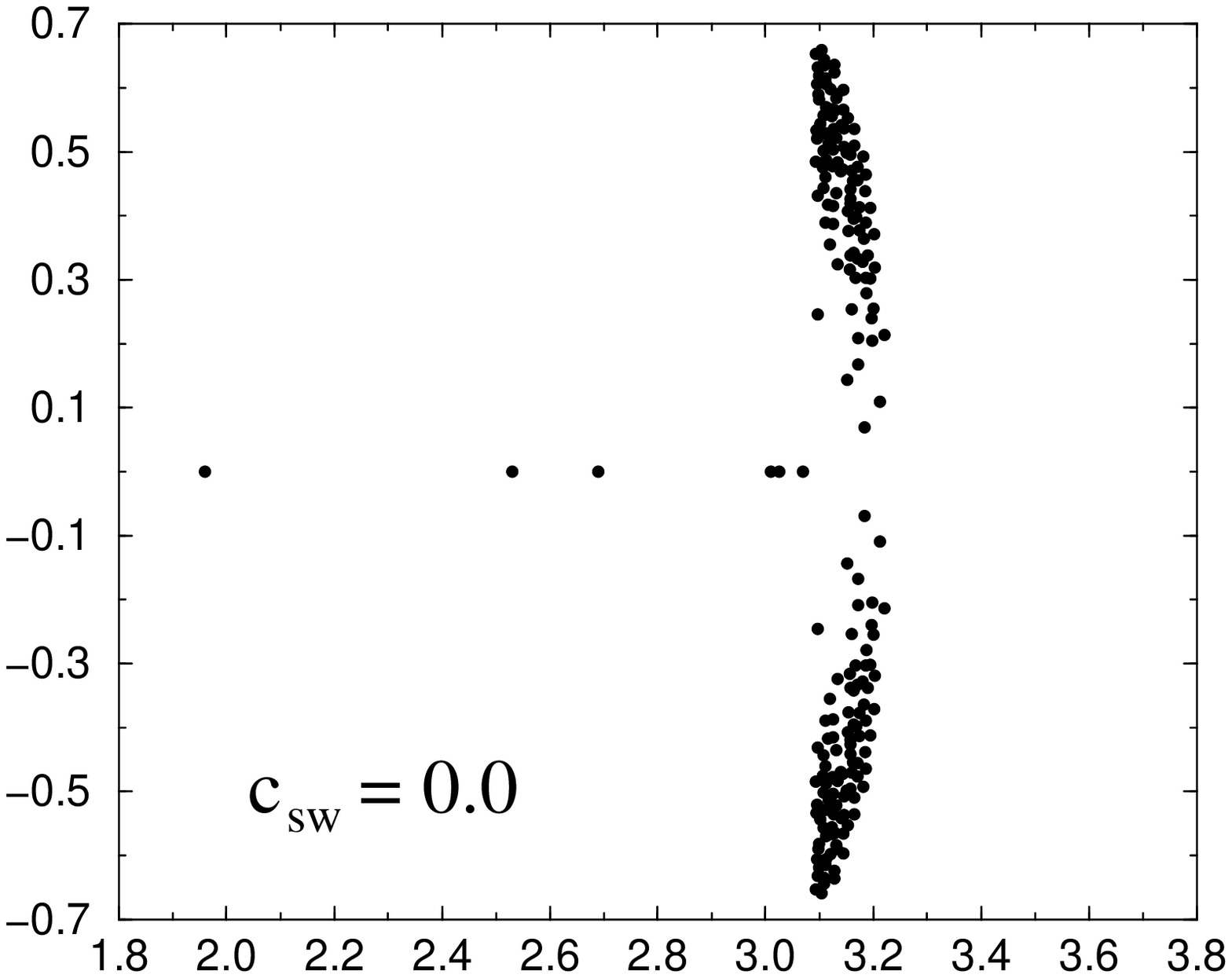} \hspace{-3mm}
\epsfysize=5.3cm\epsfbox[92 68 519 434] {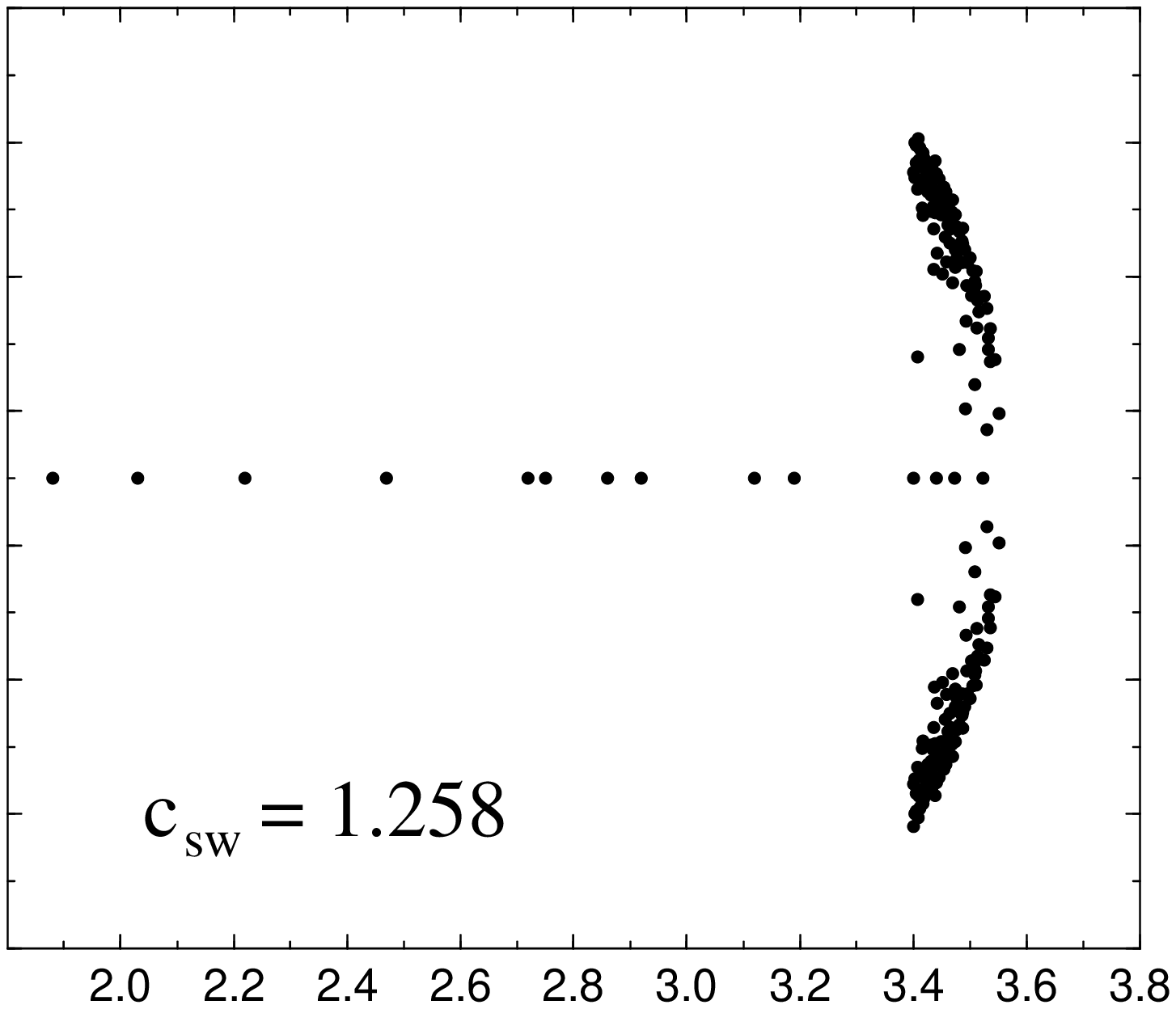} 
} 
\vspace{5mm}
\centerline{\hspace*{-1mm}
\epsfysize=5.3cm\epsfbox[56 68 519 434] {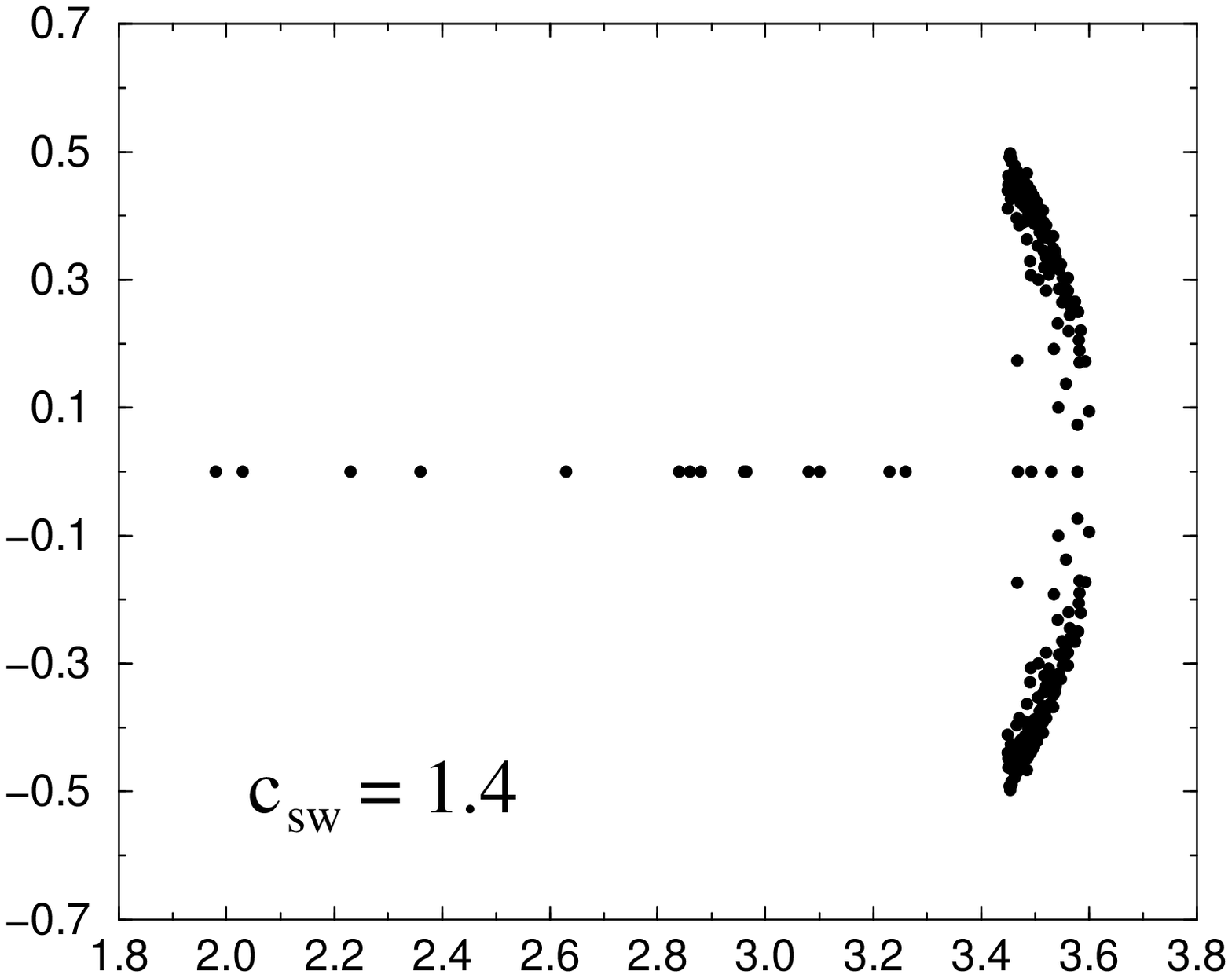} \hspace{-3mm}
\epsfysize=5.3cm\epsfbox[92 68 519 434] {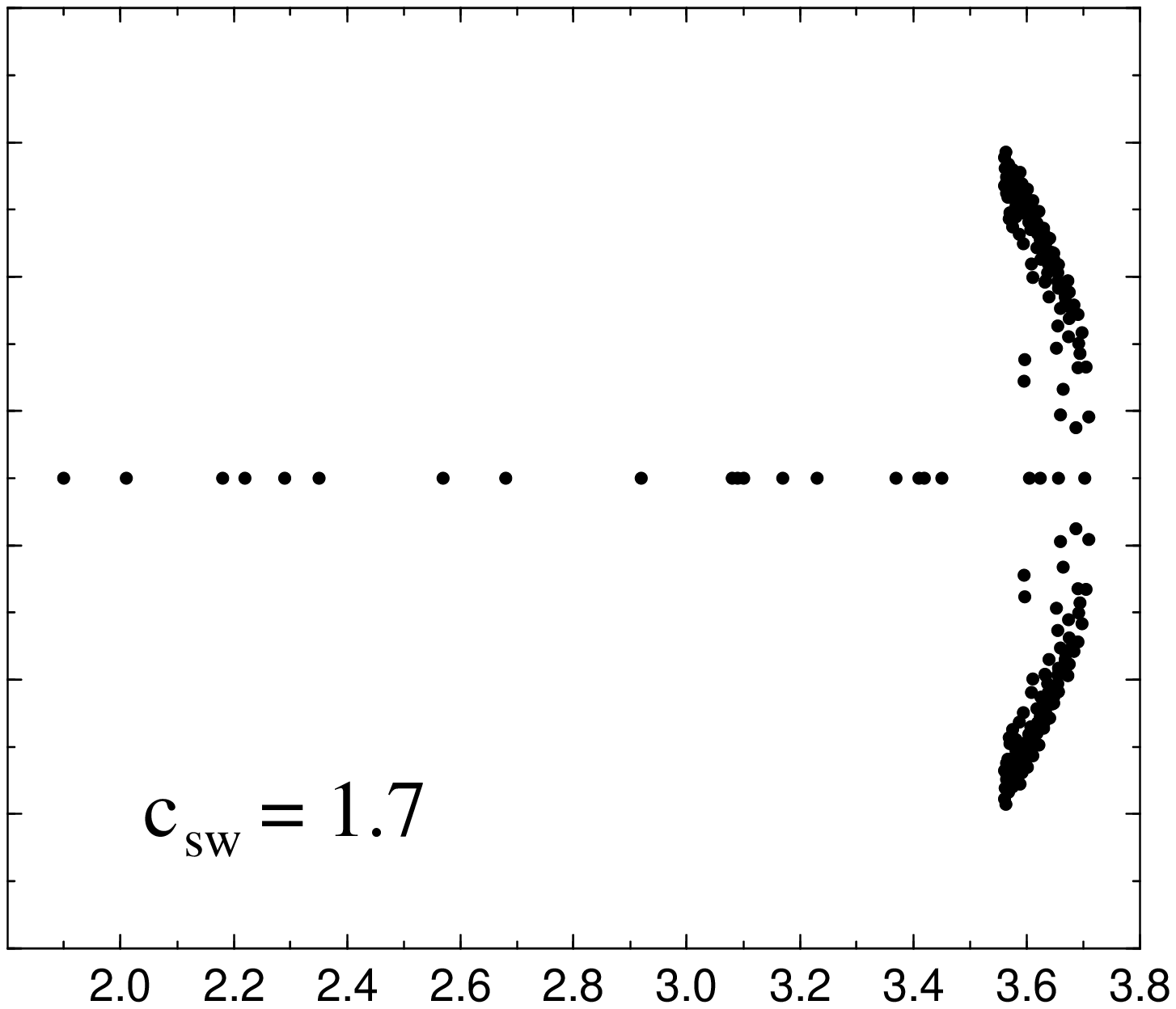}
}
\caption{{\sl Plots of $200$ complex
eigenvalues in the physical branch of the spectrum 
and all real eigenvalues larger than $1.8$ for the matrix
$K$ in a thermalized gauge 
field configuration ($12^4$ lattice, $\beta = 2.4$). 
We show the results for $c_{sw} = 0.0$
(no improvement), $c_{sw} = 1.258$ (perturbative value at $\beta = 2.4$),
for $c_{sw} = 1.4$ and for $c_{sw} = 1.7$.}
\label{hotplots}}
\end{figure}

In Fig.~\ref{hotplots} we show our results for the spectrum 
of $K$ (200 complex eigenvalues and all real eigenvalues larger than 1.8)
in a thermalized background configuration from \cite{FoGaSt97}
($12^4$ lattice, SU(2), $\beta = 2.4$). We remark that this is the thermalized 
configuration which corresponds to the cooled configuration used for the
left plot in Fig.~\ref{smoothplots} (after cooling this configuration has
$\nu = -1$). We display the spectra for $c_{sw} = 0.0$ (no improvement),
$c_{sw} = 1.258$ (perturbative value) and $c_{sw} = 1.4$ and 1.7. 
The last two
values were chosen because one expects the non-perturbative 
value of $c_{sw}$ to be larger than the perturbative 
result similar to the case of 
SU(3) where non-perturbative results for $c_{sw}$ are known \cite{nonpert}.
\\
\\
\\ 

An interesting feature of the plots is the strong 
shift of the eigenvalues in 
the physical branch of the spectrum towards larger 
real parts as $c_{sw}$ is 
increased (compare also the perturbative arguments below). This is equivalent 
to the known shift of the critical $\kappa$ when increasing $c_{sw}$ 
(compare e.g.~the first article listed in \cite{nonpert}).
We furthermore find that the eigenvalues tend to 
cluster more in a narrow 
band, although the improvement of the spectral properties (= eigenvalues
lined up along a single curve) is not as
strong as for the cooled configurations.  

For the manifestation of the index theorem on the lattice 
and also other ideas such as the modified quenched approximation
\cite{BaDuEiHoTh98}
the most important feature of the clover term is however 
the generation of additional real eigenvalues. This 
property can be seen clearly 
from the four plots in Fig.~\ref{hotplots}. In particular we found, that
the additional eigenvectors with real eigenvalues are generated in pairs of
opposite chirality. This fact can be understood by analyzing how
the spectral flow $\mu(\rho)$ for the auxiliary problem $H(\rho)$ changes
with $c_{sw}$: Increasing 
$c_{sw}$ can cause a flow line $\mu(\rho)$,
which had no zero crossing at $c_{sw} = 0$, to develop a
local maximum (or minimum) which eventually crosses zero from below (above)
as $c_{sw}$ is increased further. Two new crossings appear, one with 
positive and one with negative slope.
Thus, due to (\ref{slope}), 
this gives rise to two real eigenvalues with eigenvectors having
opposite chirality. We remark that the discussed mechanism is the only
way that additional zero crossings, i.e. real eigenvalues of $K$ ($D$)
can emerge, since the total number of zero crossings is even. This follows
from the continuity of the flow lines in $\rho$ and the simple limiting
behavior of $H(\rho)$. For $\rho \rightarrow \pm \infty$ one has 
$H(\rho) \rightarrow \pm \gamma_5 
\mbox{1\hspace{-1mm}I}_{volume \times N_{color}}$,
and the limiting matrices both have $2 \times volume 
\times N_{color}$ eigenvalues of
$+\rho$ and $2 \times volume \times N_{color}$ eigenvalues $-\rho$. 
The flow lines connect the eigenvalues of the limiting cases 
and thus only an even number of zero crossings can emerge. Hence the 
above discussed mechanism for the creation of zero crossings is
the only possibility to create new crossings, implying that the additional
real eigenvalues of $K$ ($D$) are always created in pairs of opposite
chirality.

\begin{figure}[htp]
\centerline{\hspace*{-4mm}
\epsfysize=6.5cm \epsfbox[29 47 536 460] {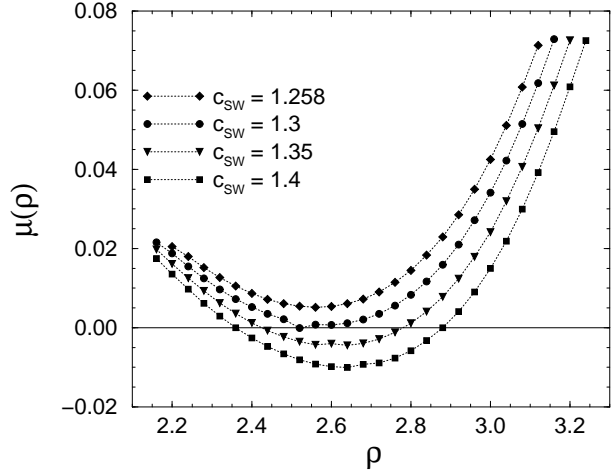}} 
\caption{{\sl Plot of the spectral flow $\mu(\rho)$ for a small eigenvalue 
of the auxiliary operator $H(\rho)$ at $c_{sw} = 1.258, 1.3, 1.35$
and $c_{sw} = 1.4$. The flow develops two zero crossings
as $c_{sw}$ is increased giving rise to two real eigenvalues of $K$ 
(or $D$)
which have eigenvectors with opposite chirality due to} 
(\protect{\ref{slope}}).
\label{prolif}}
\end{figure}

In order to demonstrate this mechanism we 
isolated the flow $\mu(\rho)$
of a single small eigenvalue of $H(\rho)$ which exhibits
the discussed phenomenon and present this plot in Fig.~\ref{prolif}.
We show how a particular flow line creates two 
new zero crossings as $c_{sw}$ is increased from the perturbative
value $c_{sw} = 1.258$ to $c_{sw} = 1.4$. We also show the flow
for the intermediate values $c_{sw} = 1.3$ and $c_{sw} = 1.35$.
For $c_{sw} = 1.258$ the flow has no crossing of zero. When increasing the 
coefficient to $c_{sw} = 1.3$ we find that 
a clear minimum has developed and the flow line is almost touching 
zero. At $c_{sw} = 1.35$ it has moved further down giving 
rise to two crossings of zero,
i.e. two additional real eigenvalues of $K$ ($D$). 
At $c_{sw} = 1.4$ the crossings of zero have moved a little and
are now at approximately 2.35 (negative slope, i.e. negative chirality) 
and at 2.9 (positive chirality). 
\\

To further study the phenomenon of proliferation of real eigenvalues when
increasing $c_{sw}$ we numerically evaluated some of the matrix
elements which occur in first and second 
order perturbation theory (in $c_{sw}$)
for eigenvalues $\lambda(c_{sw})$ of the operator $K = Q - c_{sw} C$
\begin{equation}
\lambda(c_{sw}) \; = \; \lambda(0) \; - \; 
c_{sw} {\psi_{\lambda(0)}}^\dagger C \psi_{\lambda(0)} \; + \;
c_{sw}^2 \sum_{\mu \neq \lambda(0)} \frac{\Big| 
{\psi_{\lambda(0)}}^\dagger C \psi_\mu \Big|^2}{\lambda(0) - \mu} 
\; + \; O(c_{sw}^3) \; .
\label{pert}
\end{equation}
Here $\lambda(0)$ is the eigenvalue of $K$ without improvement 
($c_{sw} =0$) and $\psi_{\lambda(0)}$ is the corresponding eigenvector.
Since $C$ is a hermitian matrix it is clear 
that the first order term is real 
and simply shifts the eigenvalues parallel to the real axis. 
Already existing
real eigenvalues remain real. This linear term seems to be the main 
contribution for the shift of the spectrum towards larger real 
parts which was already discussed for Figs.\ref{spectra}, 
\ref{smoothplots} and \ref{hotplots}.
When analyzing the second 
order term for some real eigenvalue $r$ we find 
$|\psi_r^\dagger C \psi_\mu|^2
= |\psi_r^\dagger C \psi_{\overline{\mu}}|^2$. 
Since both, $\mu$ and $\overline{\mu}$ occur in the sum for 
the second order term the imaginary parts cancel and
we find that the real eigenvalue $r$ is only 
shifted along the real axis. Again we see that existing real eigenvalues 
remain real.
For complex eigenvalues $\lambda$ we found that the second
order term leads to a shift essentially parallel to the imaginary axis and
always directed towards the real axis. Thus in second order
complex eigenvalues can 
move closer to the real axis and eventually become real eigenvalues.

To summarize this section we find that for thermalized gauge fields
the clover term leads to a proliferation 
of real eigenvalues. The above given argument, based on the spectral flow 
for the eigenvalues of the auxiliary problem, shows that the eigenvalues
are generated in pairs and the corresponding eigenvectors have opposite 
chirality. A perturbative analysis up to second order in $c_{sw}$ shows
that existing real eigenvalues remain real, but complex eigenvalues 
tend to move towards the real axis. 
\\
\\
{\bf 6. Index theorem and fermionic definition of the topological charge}
\\
The fact that the additional real eigenvalues come in pairs with opposite
chirality means that in principle they cancel each other in the lattice 
version (\ref{lasit}) of the index theorem. However it is also obvious from
Fig.~\ref{hotplots} that already for $c_{sw} = 0$ 
the separation of physical
and doubler branches is not very pronounced 
and the additional real eigenvalues
make this situation worse for $c_{sw} > 0$.
In this section we concentrate on the real 
spectrum and address the question if the separation 
of physical and doubler branches
is large enough so that one can speak of a 
probabilistic lattice manifestation
of the index theorem in a meaningful way.

\begin{figure}[htp]
\centerline{
\epsfysize=5.8cm\epsfbox[0 0 461 215] {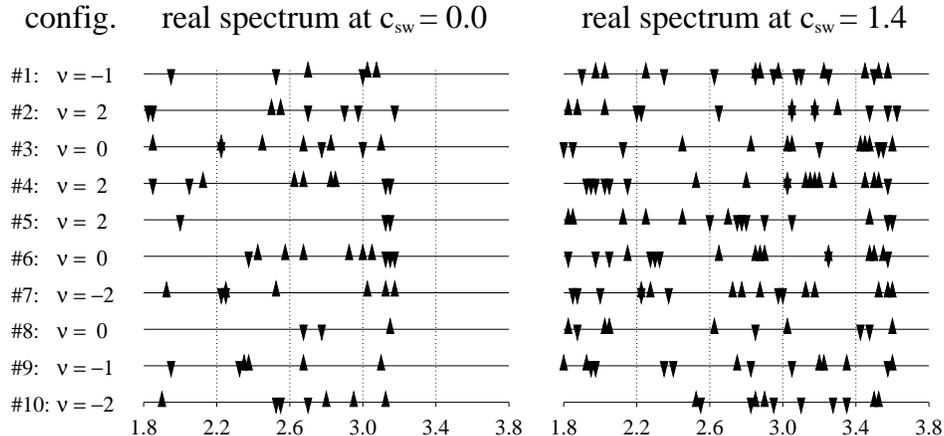}} 
\caption{{\sl Real eigenvalues larger than $1.8$ for 
$10$ thermalized gauge field 
configurations. The value of $\nu$ given on the right hand side of the plot 
was assigned after cooling} \protect{\cite{FoGaSt97}}. {\sl
An upward pointing triangle is the symbol for an eigenvalue with positive
chirality while a downward pointing triangle indicates negative chirality.}
\label{realplots}}
\end{figure}

Fig.~\ref{realplots} shows our results for 10 thermalized 
configurations which for further reference were numbered (\#1 - \#10). 
We also quote the topological charges that were assigned to these 
configurations after cooling in \cite{FoGaSt97}. The figure 
shows the real spectra for $c_{sw} = 0$ and $c_{sw} = 1.4$. 
For both cases we computed all real eigenvalues larger than 1.8 
in order to see at least a few of the real eigenvalues in the doubler 
branch. In the plot we also encode the chirality of 
the corresponding eigenstates. An upward pointing triangle means 
positive and a downward pointing triangle indicates negative chirality.

Let's first discuss the spectra for the case without clover term. From the 
arbitrarily chosen sample we show here it is obvious, that the real 
eigenvalues are rather evenly distributed and only for 
configuration \# 5 one can speak of a reasonably separated 
physical branch. For the other configurations it is unclear where to 
set the threshold for the physical modes. 

We conclude that for this particular setting (SU(2), $12^4$, $\beta$ = 2.4)
the naive approach to a probabilistic lattice interpretation of the 
Atiyah-Singer index theorem by simply counting the real eigenvalues in
the physical branch seems problematic.

In this context it is interesting to remark, 
that Narayanan and Vranas \cite{NaVr97}
obtained a surprising result using the fermionic overlap
definition of the topological charge
and averaging over large samples of quenched gauge field configurations
(also for SU(2), $12^4$, $\beta = 2.4$). Narayanan and Vranas take into 
account all real eigenvalues larger than 3 and discard the smaller ones 
as doublers. From our plots it is clear that even if one is willing 
to define 3 to be the boundary between physical
modes and doublers this definition disagrees in several cases 
(\# 2, \# 6, \# 7, \# 8, \# 10) with the cooling results and is not 
justified on a one by one basis. 
However after sampling 140 configurations \cite{NaVr97},
the result for the distribution of the topological sectors 
\cite{FoGaSt97} from improved cooling is reproduced with good 
accuracy. This comparison of the averaged result with the analysis of single
configurations indicates that there might be a mechanism which averages the
results from counting the eigenvalues such that the distribution of the
topological charge matches the outcome of the improved cooling analysis.

As already discussed in Section 5, the clover term does not improve  
the situation for
the lattice index theorem. For a larger sample this is now
demonstrated on the right hand side of Fig.~\ref{realplots} where we 
show the 10 real spectra now at Sheikholeslami-Wohlert 
coefficient $c_{sw} = 1.4$. Although 
the additional real eigenvalues come in pairs of opposite
chirality and thus in principle
cancel in (\ref{lasit}), from a technical point of view
the extra real eigenvalues are very unpleasant since the idea of dividing 
the eigenvalues into physical and doubler modes becomes even more 
unrealistic than at $c_{sw} = 0$. However, also for $c_{sw} > 0$ a
phenomenon similar to the $c_{sw} = 0$ case might occur, which would give
the correct distribution of the topological charge when averaging large
samples. It would be interesting to study and understand such a possibility 
in more detail. 
\\
\\
{\bf 7. Discussion}
\\
In this article we have studied the spectrum of the lattice Dirac operator
and investigated the interplay between topological charge and spectral
properties. The effects of the $O(a)$-improving clover term were analyzed.

For relatively smooth toy configurations 
(constant plaquette fields + small fluctuations) 
on $4^4$ lattices we find 
that the clover term leaves the physical branch of the spectrum rather 
unchanged while the bulk of the eigenvalues is slightly deformed. For 
the smooth toy configurations the lattice realization of the Atiyah-Singer
index theorem is not affected by the clover term.

When analyzing smooth configurations on larger lattices obtained 
using the improved cooling method in \cite{FoGaSt97}, we even find 
improvement of the spectral properties when adding the clover term, 
i.e.~the eigenvalues at the physical edge of the spectrum
allign along an ellipse and the pseudoscalar 
matrix elements are close to $\pm 1$.

For the thermalized configurations ($12^4, \beta = 2.4$)
a naive attempt to interpret the real spectrum in terms of  
a probabilistic manifestation of the index theorem by simply counting the
real eigenvalues in the physical branch is problematic.
We observe, that there is no reasonable separation of physical modes
and the doubler branches. The real eigenvalues in the physical branch
which have to be taken into account for the lattice version of the
index theorem cannot be identified reliably for this setting.
We remark however, that summing over large samples seems to average out the 
error in the number of real eigenvalues in the physical branch and the 
distribution of the topological charge agrees with the result from improved 
cooling \cite{NaVr97}. It would be interesting to understand
this mechanism in more detail. 
 
Adding the clover term makes the separation of physical 
eigenmodes and doublers even more difficult by creating
new real eigenvalues. 
Based on an argument using zero crossings of the spectral flow of the 
auxiliary operator $H(\rho)$ we showed that the additional real 
eigenvalues come in pairs and the corresponding eigenvectors have 
opposite chirality. A perturbative argument indicates that existing real 
eigenvalues are not destroyed by the clover term.

Certainly there is hope, that this situation improves as one goes
over to larger lattices and higher $\beta$. Such an improvement of the
spectral properties, i.e.~a clearer separation of physical modes and doublers,
leading to a simple probabilistic realization of the index
theorem as one gets closer to the continuum limit 
was observed for QED$_2$. Only further analysis can settle the 
question if a similar scenario holds for SU(N) in 4 dimensions. 
Also new approaches such as the interpolation idea
\cite{He98} or perfect actions \cite{perfact}
or other actions \cite{Ne98,Ch98} which obey the Ginsparg-Wilson relation 
\cite{GiWi82} and are known to obey the index theorem \cite{Lu98b}
would be interesting to study numerically in 4 dimensions.
\\
\\
{\bf Acknowledgements:} We would like to thank Philippe de Forcrand,
Ion-Olimpiu Stamatescu and in particular Margarita Garc\'{\i}a P\'erez
for letting us use their configurations from \cite{FoGaSt97} and providing
us also with the cooled data thus allowing for a comparison on a one by one
basis. We also acknowledge interesting discussions and help from 
Christian Lang, Stefan Sint and Peter Weisz.
\\


\begin{thebibliography}{1234567}
\newcommand{\bibi}[1]{\bibitem{#1}}
\newcommand{\authors}[1]{#1, }
\newcommand{\journal}[1]{#1}
\newcommand{\volume}[1]{#1}
\newcommand{\myyear}[1]{(#1)}
\newcommand{\page}[1]{#1}
\newcommand{\mytitle}[1]{}
\newcommand{\keywords}[1]{}
\newcommand{\kw}[1]{}
%
\bibi{AtSi68}
\authors{M. Atiyah and I.M. Singer}
\journal{Ann. \ Math.}
\volume{87} \myyear{1968} \page{596},
\journal{Ann. \ Math.}
\volume{93} \myyear{1971} \page{139}.
%
\bibi{ItIwYo87}
\authors{S. Itoh, Y. Iwasaki and T. Yoshi\'e}
\journal{Phys. Rev.}
\volume{D36} \myyear{1987} \page{527}, 
\journal{Phys. Lett.}
\volume{184B} \myyear{1987} \page{375}.
%
\bibi{SmVi87}
\authors{J. Smit and J.C. Vink}
\journal{Nucl. Phys.}
\volume{B286} \myyear{1987} \page{485}.
%
\bibi{NaNe95}
\authors{R. Narayanan and H. Neuberger}
\journal{Nucl. Phys.}
\volume{B412} \myyear{1994} \page{574},
\journal{Nucl. Phys.}
\volume{B443} \myyear{1995} \page{305}. 
%
\bibi{smooth}
\authors{R. Setoodeh, C.T.H. Davies and I.M. Barbour}
\journal{Phys. Lett.}
\volume{213B} \myyear{1988} \page{195};
\authors{M.L. Laursen, J. Smit and J.C. Vink}
\journal{Nucl. Phys.}
\volume{B343} \myyear{1990} \page{522};
\authors{M. Garc\'{\i}a P\'erez, A. Gonz\'alez-Arroyo, 
A. Montero and C. Pena}
\journal{Nucl.~Phys.~Proc.~Suppl.}
\volume{63} \myyear{1998} \page{501}.
%
\bibi{Ne97}
\authors{J.W. Negele}
\journal{Reports hep-lat/9709129, hep-lat/9804017}; 
\authors{T.L. Ivanenko and J.W. Negele}
\journal{Nucl.~Phys.~Proc.~Suppl.}
\volume{63} \myyear{1998} \page{504}.
%
\bibi{GaHi97}
\authors{C. Gattringer and I. Hip}
\journal{Report hep-lat/9712015, to appear in Nucl.~Phys.~B}.
%
\bibi{EdHeNaSi98}
\authors{R.G. Edwards, U.M. Heller and R. Narayanan}
\journal{Nucl. Phys.}
\volume{B522} \myyear{1998} \page{285}.
%
\bibi{qed2}
\authors{J.C. Vink}
\journal{Nucl. \ Phys.}
\volume{B307} \myyear{1988} \page{549};
\authors{R. Narayanan, H. Neuberger and P. Vranas}
\journal{Phys. Lett.}
\volume{B353} \myyear{1995} \page{507};
\authors{W. Bardeen, A. Duncan, E. Eichten and H. Thacker}
\journal{Phys. Rev.}
\volume{D57} \myyear{1998} \page{3890}.
%
\bibi{He98}
\authors{P. Hernandez}
\journal{Report hep-lat/9801035}. 
%
\bibi{GaHiLa97}
\authors{C.R. Gattringer, I. Hip and C.B. Lang}
\journal{Nucl.~Phys.}
\volume{B508} \myyear{1997} \page{329},
\journal{Nucl.~Phys.~Proc.~Suppl.}
\volume{63} \myyear{1998} \page{498}.
%
\bibi{NaVr97} 
\authors{R. Narayanan and P. Vranas}
\journal{Nucl. Phys.}
\volume{B506} \myyear{1997} \page{373}.
%
\bibi{NaSi98}
\authors{R. Narayanan and R.L. Singleton Jr.}
\journal{Nucl. Phys. Proc. Suppl.}
\volume{63} \myyear{1998} \page{555}.
%
\bibi{Naetal98}
\authors{R.G. Edwards, U.M. Heller and R. Narayanan}
\journal{Nucl. Phys.}
\volume{B518} \myyear{1998} \page{319},
\journal{Nucl. Phys.}
\volume{B535} \myyear{1998} \page{403}.
%
\bibi{BaDuEiHoTh98}
\authors{W. Bardeen, A. Duncan, E. Eichten, G. Hockney and H. Thacker}
\journal{Phys. Rev.}
\volume{D57} \myyear{1998} \page{1633}.
%
\bibi{JaLiSiSm97}
\authors{K. Jansen, C. Liu, H. Simma and D. Smith}
\journal{Nucl. Phys. Proc. Suppl.}
\volume{53} \myyear{1997} \page{262}.
%
\bibi{SmSiTe98}
\authors{D. Smith, H. Simma and M. Teper}
\journal{Nucl. Phys. Proc. Suppl.}
\volume{63} \myyear{1998} \page{558};  
\authors{H. Simma and D. Smith}
\journal{Report hep-lat/9801025}.
%
\bibi{FoGaSt97}
\authors{P. de Forcrand, M. Garc\'{\i}a P\'erez and I.-O. Stamatescu}
\journal{Nucl. Phys.} 
\volume{B499} \myyear{1997} \page{409}.
%
\bibi{impcool2}
\authors{P. de Forcrand, M. Garc\'{\i}a P\'erez and I.-O. Stamatescu}
\journal{Nucl.~Phys.~Proc.~Suppl.}
\volume{47} \myyear{1996} \page{777};
\authors{P. de Forcrand, M. Garc\'{\i}a P\'erez, James E. Hettrick
and I.-O. Stamatescu}
\journal{Nucl.~Phys.~Proc.~Suppl.}
\volume{63} \myyear{1998} \page{549},
\journal{Report hep-lat/9802017}.
%
\bibi{GaGoSnBa94}
\authors{M. Garc\'{\i}a P\'erez, A. Gonz\'alez-Arroyo, 
J. Snippe and P. van Baal}
\journal{Nucl. Phys.} 
\volume{B413} \myyear{1994} \page{535}.
%
\bibi{Lu98}
\authors{M. L\"uscher}
\journal{{\sl Advanced Lattice QCD} (Les Houches 1997),
Report hep-lat/9802029}.
%
\bibi{qed2perf}
\authors{F. Farchioni and V. Laliena}
\journal{Phys. Rev.}
\volume{D58}:054501 \myyear{1998} ;
\authors{F. Farchioni, C.B. Lang and M. Wohlgenannt}
\journal{Phys. Lett} 
\volume{B433} \myyear{1998} \page{377}.
%
\bibi{perfact}
\authors{P. Hasenfratz, V. Laliena and F. Niedermayer}
\journal{Phys. Lett.}
\volume{B427} \myyear{1998} \page{125};
\authors{P. Hasenfratz}
\journal{Nucl. Phys.}
\volume{B525} \myyear{1998} \page{401}.
%
\bibi{LuSiSoWe97}
\authors{M. L\"uscher, S. Sint, R. Sommer and P. Weisz}
\journal{Nucl. Phys.}
\volume{B478} \myyear{1996} \page{365}.
%
\bibi{ShWo85}
\authors{B. Sheikholeslami and R. Wohlert}
\journal{Nucl. Phys.}
\volume{B259} \myyear{1985} \page{572}.
%
\bibi{LuWe96}
\authors{M. L\"uscher and P. Weisz}
\journal{Nucl. Phys.}
\volume{B479} \myyear{1996} \page{429}.
%
\bibi{MoMu94}
\authors{I. Montvay and G. M\"unster} {\sl Quantum Fields on a Lattice},
Cambridge University Press, Cambridge 1994.
%
\bibi{WeCh79}
\authors{D.H. Weingarten and J.L. Challifour}
\journal{Ann. Phys.}
\volume{123} \myyear{1979} \page{61}.
%
\bibi{So92}
\authors{D.C. Sorensen}
\journal{SIAM J. Matrix Anal. Appl.}
\volume{13} \myyear{1992} \page{357}. 
%
\bibi{nonpert}
\authors{M. L\"uscher, S. Sint, R. Sommer, P. Weisz and U. Wolff}
\journal{Nucl. Phys.} 
\volume{B491} \myyear{1997} \page{323};
\authors{R.G. Edwards, U.M. Heller and T.R. Klassen}
\journal{Nucl.~Phys.~Proc.~Suppl.}
\volume{63} \myyear{1998} \page{847};
\authors{K. Jansen and R. Sommer}
\journal{Nucl.~Phys.~Proc.~Suppl.}
\volume{63} \myyear{1998} \page{853}.
%
\bibi{Ne98}
\authors{H. Neuberger}
\journal{Phys. Lett.}
\volume{B417} \myyear{1998} \page{141},
\journal{Phys. Lett.}
\volume{B427} \myyear{1998} \page{353},
\journal{Report hep-lat/9806025}.
%
\bibi{Ch98}
\authors{T.-W. Chiu}
\journal{Phys. Rev.}
\volume{D58}:074511 \myyear{1998};
\authors{T.-W. Chiu and S.V. Zenkin}
\journal{hep-lat/9806019}.
%
\bibi{GiWi82}
\authors{P.H. Ginsparg and K.G. Wilson}
\journal{Phys. Rev.}
\volume{D25} \myyear{1982} \page{2649}.
%
\bibi{Lu98b}
\authors{M. L\"uscher}
\journal{Phys. Lett.}
\volume{B428} \myyear{1998} \page{342}.

\end{thebibliography}
\end{document}